# Towards the Patterns of Hard CSPs with Association Rule Mining

Chendong Li
*Texas Tech University*
chendong.lcd@gmail.com

## Abstract

*The hardness of finite domain Constraint Satisfaction Problems (CSPs) is a very important research area in Constraint Programming (CP) community. However, this problem has not yet attracted much attention from the researchers in the association rule mining community. As a popular data mining technique, association rule mining has an extremely wide application area and it has already been successfully applied to many interdisciplines. In this paper, we study the association rule mining techniques and propose a cascaded approach to extract the interesting patterns of the hard CSPs. As far as we know, this problem is investigated with the data mining techniques for the first time. Specifically, we generate the random CSPs and collect their characteristics by solving all the CSP instances, and then apply the data mining techniques on the data set and further to discover the interesting patterns of the hardness of the randomly generated CSPs.*

## 1. Introduction

In the Constraint Programming (CP) community, the randomly generated CSP instances are usually used as the empirical criteria to evaluate and benchmark the novel proposed algorithms on solving the finite domain Constraint Satisfaction Problems (CSPs).

### 1.1. The CSPs hardness

The hard CSP instances are highly evaluated since they can be used to recognize whether or not a CSP solver is fast than others with the significant difference on the running time. These hard CSP instances are highly desired in the CP community. On the one hand, they are of great significance for the experimental evaluation of CSP solvers, and on the other hand, the hard CSPs could contribute to the evaluation criteria for the theoretical computer science. Generating hard CSPs is of great important problems. These random CSP instances may be further employed to help analyzing the performance of the new algorithms in the average cases. Starting from Cheesman and Kanefsky et al. (1991) [7], a lot of researchers in the CP and SAT community begin to investigate on the CSPs hardness (for details, please refer to the related work).

The previous researches on the hardness of CSPs mainly concentrated on the general CSPs — no constraints types were specified. Usually the hard CSPs discovered in the randomly generated instances are defined over the parameters such as the number of variables $N$, domain size $D$, density of the constraint network associated with tightness of constraints, which formulate the finite constraint set $C$. Since functional constraint are recognized as a very common class of constraints [12, 14], more and more researchers put many efforts to study functional constraints, such as [2, 19] especially the bi-functional constraint [14, 20], which is considered as the tractable constraint class. Because the functional constraints are easier to solve than the general constraints, the task of finding the hard CSPs with functional constraints become extremely challenging.

### 1.2. Association rule mining

Association rule mining is a very widely utilized data mining technique that has already been successfully applied to market basket analysis (such as determining customer buying patterns), intrusion detection, financial profiles [34], bioinformatics [22], web-based applications [25], privacy information [35], and so on. Most researchers in the association rule mining community put much a lot of efforts on exploring the new, fast and efficient association rule mining algorithms, such as [21, 23, 24, 26, 27, 36]. This paper contributes to the new application of association rule on extracting the patterns of hard CSP instances, so that it can provide useful recommendations for researchers who want to generate the hard CSPs to benchmark their new algorithms.

Association rule mining has already been successfully applied to determine customer buying patterns so that the supermarkets can sell out more products to customers according to the extracted patterns. Similarly, association rule can be used to discover the patterns of the hard CSPs by applying association rule mining techniques on the important characteristics of a set of CSP instances.

The rest of the paper is organized as follows. Section 2 presents the related work. Background on CSPs and association rules are given in Section 3. Section 4 describes the proposed cascaded approach. Section 5 details the experiments and the results. Section 6 gives a brief discussion and remarks on the proposed method. Section 7 concludes the paper and the directions for future work are presented in Section 8.

## 2. Related work

In this part, since no relevant work on the association rule mining with the CSP hardness has been done, we mainly focus on the related work on the hard CSPs.

The analytical analysis on the exponentially long tail of CSP hardness distributions was provided by Hogg and Williams (1994) in [10]. Mitchell (1994) [16] shows the empirical results that with the increasing number of samples recorded, the mean of running time, standard deviation, and maximum time are also increase correspondingly. The instances from very natural CSP models have been studied by Prosser (1996) in [8]. Selman and Kirkpatrick (1996) study this problem and further analyzed the distributions of both satisfiable and unsatisfiable instances in [13]. The empirical facts that neither the hardness of randomly generated CSPs nor 3-coloring problems is normally distributed was presented by Kwan (1996) in [11]. Frost and Rish et al. (1997) in [17] provides the empirical evidence that the randomly generate CSPs at the 50% satisfiable point, satiafiable problems can be modeled by Weibull distribution and unsatifiable instances by the lognormal distribution. In [5] Mitchell (1998) proves the exponential lower bounds for the running time of many CSP solving techniques.

In [4], Mitchell (2002) shows that for a random CSP of which the constraints are loose or reasonably tight, the majority of unsatisfiable CSP families have a structural property that make it holds that the exponential size of the unsatisfiability proofs must exist in a certain resolution-like system. In [1], Mitchell (2003) shows that each CNF formulas family that is hard for the propositional resolution corresponds to the CSP instances that are hard for most of the standard CSP algorithms. In [4] Xu and Li (2004) study the solution structures of the random CSPs and random k-SAT problems and utilized the concept of average similarity degree to characterize the similarity of the solutions between random CSPs and random k-SAT. In [37], Prcovic (2005) considered tree search procedures that filter the domains dynamically by maintaining the local consistency like Forward-Checking and MAC. A new method based on the expander graphs was proposed to generate the hard k-SAT and CSP instances in [3] (2008). In [15], O'Donnell and Wu (2009) investigated on the conditional hardness for the 3-CSPs that can be satisfied.

## 3. Preliminaries

In this section, we introduce the basic concepts of CSP and association rule mining used in this paper.

### 3.1. Constraint Satisfaction Problems

**Definition 1** (Constraint Satisfaction Problems) A finite domain Constraint Satisfaction Problem (CSP) is defined as the triple (*N*, *D*, *C*) that consists of the following parts:

(i) a finite set of variables $N = \{1,2,...,n\}$,

(ii) a discrete and finite set of Domains $D = \{D_1, D_2, ..., D_n\}$, where $D_i$ is the domain of viable $i$, $i \in N$, i.e., $1 \leq i \leq n$.

(iii) a set of constraints *C*, each of which is a relation among the variables in *N*.

A binary Constraint Satisfaction Problem (BCSP) is a CSP that satisfies the above (i) and (ii); the only difference is the constraint set *C*. For a BCSP, the constraint is defined as the binary relation between any two variables in the variable set *N*. A constraint over two variables $i$ and $j$ is denoted by $c_{ij}$. The constraint $c_{ij}$ is functional on $j$ if for any $a \in D_i$ there exists at most one $b \in D_j$ such that the triple $(a, b)$ satisfies the constraint $c_{ij}$. A functional constraint is a constraint $c_{ij}$ that is either functional on $j$ or $i$. If the $c_{ij}$ is both functional on $j$ and $i$, it is called the bi-functional constraint. A special case of functional constraints are equations, which are ubiquitous in the Constraint Logic Programming ($\mathcal{CLP}$). The typical functional constraint in arithmetic is a binary linear equation like $5x = 3 + 2y$, which is both functional on $x$ and on $y$.

For a CSP, the tightness of the constraint is defined as the percentage of the allowed tuples over $d^2$, where $d$ is the domain size of the variables.

The process of solving a CSP is either to find an assignment that satisfies all the constraints or to prove that no such assignment exists for the given problem. In general, solving a CSP is NP-complete. Although it is known that Constraint Satisfaction Problems (CSP) are NP-complete, only a small percentage of the whole CSP set are hard to solve with the current backtracking algorithms. For example, based on our computational results, only 2.57% of the total CSPs have the running time longer than 5.0 seconds.

Currently there are mainly two solving techniques: 1) the local repair methods and 2) the integrated methods of the tree search and domain filtering algorithms. The first one can handle the CPSs whose solutions distribute in search tree uniformly while the second method can deal with the over-constrained CSPs [37]. If a CSP instance cannot be solved by the above two solving techniques efficiently, then it is usually classified as the hard CSP.

In the following paper, we place ourselves in the context of binary CSP, which is considered as one of the most important CSP classes. One thing needs to mention is that when attempting to generate the hard binary CSPs, people usually adjust the density of the constraint network and constraints tightness to make the CSPs as hard as possible, instead of keeping on increasing the number of variables and domain size.

### 3.2. Association rule mining

**Definition 2** (Association rule) According to Agrawal and Srikant [32], the problem of association rule mining is defined as follows: Let $IS = \{i_1, i_2, ..., i_n\}$ be the itemset and $ST = \{t_1, t_2, ..., t_m\}$ be a set of transactions, which is known as the database. Each transaction TR (with a unique transaction ID) in ST is a set of items such that $TR \subseteq IS$. An association rule is defined as an implication of the form $X \Rightarrow Y$, where $X \subseteq IS$, $Y \subseteq IS$ and $X \cap Y = \Phi$.

The association rule $X \Rightarrow Y$ holds in the set of transactions ST with the support $s$ and confidence $c$ if $s$ % of transactions in ST contain $X \cup Y$ and $c$ % of transactions in ST that contain X also contain Y. These measurements of support and confidence were original proposed by Agrawal and Srikant in [32] to evaluate the interestingness of the association rules. Specifically, for a specified association rule $X=a \Rightarrow Y=b$ the confidence can be expressed as

$$Confidence(X = a \Rightarrow Y = b) = \frac{P(X = a, Y = b)}{P(X = a)}$$
$$= P(Y=b | X=a)$$

Similarly the support of the specified rule can be expressed as

Support $(X=a \Rightarrow Y=b) = P(X=a, Y=b)$

Based on this idea, the notation of the Agrawal and Srikant in [32] can be expressed in the following way: Given the respect minimum thresholds on confidence and support, say $min\_c$, $min\_s \in [0, 1]$ the specified association rule holds in the set of rules R if the following condition holds.

Confidence $(X=a \Rightarrow Y=b) \geq min\_c \land$ Support$(X=a \Rightarrow Y=b) \geq min\_s$

The association rule $X \Rightarrow Y$ can be interpreted as "IF X THEN Y". For example, the following published rule R0 can be interpreted as: if someone is at the education level of bachelors, then he/she has a year income of less than 50K over the specified threshold of support and confidence.

R0) education = bachelor $\Rightarrow$ income <= 50K

### 3.3. Interestingness measures of association rules

Measuring the interestingness of the discovered association rules is a very active and significant research area in data mining. However, as of today there is no widespread agreement on the formal definition of interestingness in this context [28], although numerous experiments have been conducted on this subject and many different criteria and measures based on probability have been studied in order to select interesting rules from all the possible rules. Among these measures of significance and interestingness, the minimum thresholds on support and confidence are the most commonly accepted criteria.

Probability based interestingness measures and the evaluations of association rules have been systematically studied by the data mining community. A recent overview of the relevant measures is presented in [28]. In the following, we briefly review four important measures.

**Definition 3** (Support) Support [29], known as the frequency constraint, is defined on the itemsets to represent the proportion of the transactions that contain a specified itemset. It measures the significance of an itemset.

Support $(X) = P(X)$

**Definition 4** (Confidence) Confidence [29] is the probability of viewing Y in the rule $X \Rightarrow Y$ under the condition that the transactions also contain X.

Confidence $(X \Rightarrow Y) = P(Y|X)$

Support is used to find frequent itemsets while confidence is to generate rules from the frequent itemsets. Besides confidence and support, conviction and lift are also the important measures.

**Definition 5** (Conviction) Conviction [30] is the comparison of the probability that X occurs without Y if they were statistically dependent with the actual frequency of the occurrence of X without Y. It is monotone in both confidence and lift.

$$Conviction(X \Rightarrow Y) = \frac{P(X)P(\neg Y)}{P(X \neg Y)}$$

**Definition 6** (Lift) Lift [30] measures how many times X and Y occur simultaneously more often than the anticipated frequency if they are statistically independent.

$$Lift(X \Rightarrow Y) = \frac{P(XY)}{P(X)P(Y)}$$

In our experiments, we choose to use lift as the main criteria to measure the interestingness of the generated association rules by applying the Apriori algorithm.

## 4. The proposed methods

In this section, we formulate the problem of the hardness of randomly generated CSPs as the association rule mining problem. Moreover, we propose a cascaded approach to extract the interesting CSP hardness patterns with the association rule mining techniques. For the rest part, we start with the Apriori algorithm and then introduce the rule deduction. On the basis of these two techniques, we present our cascaded approach in details.

### 4.1. The Apriori algorithm

Apriori algorithm [29, 32], a classic algorithm for association rules learning, is designed to manipulate on the databases of transactions. It employs the "bottom up" approach where frequent subsets augment one item each time, which is known as the frequent itemset generation or candidate generation. It generates candidate itemsets of length k from itemsets of length k – 1 (as shown in line 3 of Figure 1) and then prunes the candidates which have an infrequent sub pattern (lines 4 and 5 of Figure 1). It utilizes a tree structure to count frequent itemsets and uses downward closure to prune unnecessary branches. Then the frequent itemsets are tested against the data. This algorithm terminates if no further augmentation can be made. The pseudo code of finding the frequent itemsets algorithm is given in Figure 1.

Two parameters are involved in the Apriori algorithm: the minimum support ($min\_s$) used for generating frequent itemsets and the other is the minimum confidence used for rule derivation.

**Input**: Database of transactions ***D***; threshold of minimum support, $min\_s$.
**Output**: frequent itemsets in ***D***.
**Pseudo-code**:
  // *Initialization*
1) $L_1$ = frequent items;
2) ***for*** ($k = 2; L_{l-1} \neq \phi; l++$) {
    // Step 1: candidate generation
3)    $C_l$ = generate ($L_{l-1}$, $min\_s$);
    // Step 2: *Pruning*
4)    ***for*** each transaction $tr \in D$ {
5)       $C_{tr}$ = subset ($C_l$, $tr$);
6)       ***for*** each candidate $c \in C_{tr}$
7)          c_count++;
    }
8)    $L_l = \{c \in C_l \mid c\_count \geq min\_s\}$;
  }
9) ***return*** $\cup_l L_l$;

**Figure 1. The Apriori algorithm of finding the frequent itemsets**

### 4.2. Rule deduction scheme

Viewed from the perspective of the deduction methods in propositional logic, a rule $X \Rightarrow Y$ is semantically entailed from a set of rules if every dataset where all the rules hold must also satisfy $X \Rightarrow Y$ [31]. Syntactically, the rule $X \Rightarrow Y$ holds if and only if $X \Rightarrow Y$ is derivable from all the rules by applying the Armstrong's Axioms [33]:

  1) Reflexivity. $X \Rightarrow X$
  2) Transitivity. If $X \Rightarrow Y$ and $Y \Rightarrow Z$ then $X \Rightarrow Z$
  3) Augmentation. If $X \Rightarrow Y$ and $Z \Rightarrow W$ then $X, Z \Rightarrow Y, W$ (Except these three axioms, there are additional theorems such as decomposition theorem.)

This kind of mechanism is called rule deduction in the following part. Based on the rule deduction, we can obtain much more meaningful information. For example, given rule R1 and R2, we can obtain rule R3 by simply applying the transitivity axioms. Furthermore, rule R4 can be obtained by applying the augmentation axiom on rule R2 and R3. (Note that the left hand of the rule R2 and R3 are the same, which means X = Z according to the Armstrong's Axioms. Therefore, to simplify the rule, we only need to keep X or Z, which is tightness = 0.620)

R1) Classification = Hard ⇒ Satisfiablity = No
R2) Tightness = 0.620 ⇒ Classification = Hard
R3) Tightness = 0.620 ⇒ Satisfiablity = No
R4) Tightness = 0.620 ⇒ Classification = Hard, Satisfiablity = No

Since applying the reflexivity axiom on any rules cannot provide any new information, we mainly consider using the transitivity axiom and augmentation axiom. Therefore, the task of finding the interesting rules from a set of association rules becomes computing the minimum closure of the set of rules.

### 4.3. The cascaded approach

In order to extract the CSPs hardness patterns effectively, we propose a cascaded approach to discover the interesting patterns of the hard CSPs as shown in Figure 2.

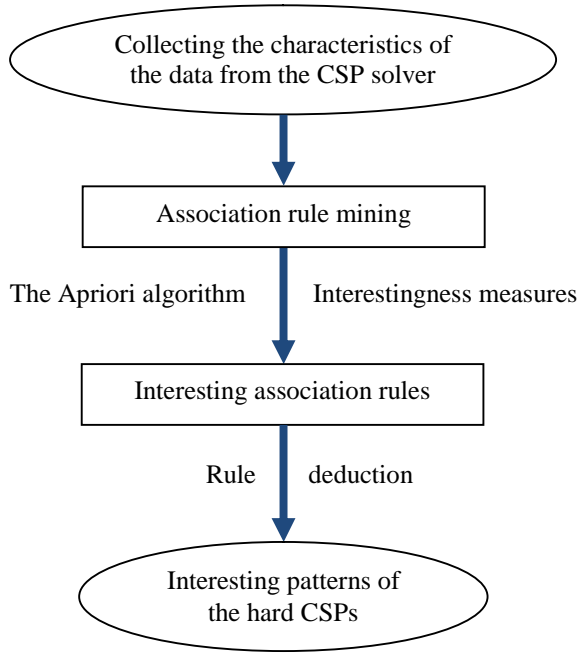

**Figure 2. The proposed approach**

In our approach, we first generate the CSP instances randomly and then solve all the instances with a general CSP solver to collect the significant characteristics of the generated CSPs. Based on these characteristics, we construct the interesting data set. And then, we apply the association rule mining techniques, which are mainly based on the Apriori algorithm. After obtaining the interesting association rules from the mining step, we further apply the rule deduction to the set of all the interesting rules to extract the interesting patterns of hard CSPs.

## 5. Experiments

To test the proposed approach, we conducted a series of experiments. As no publicly benchmarked CSPs with functional constraints are available, we generate the random CSPs. The types of constraints in the generated CSPs include functional constraints, bi-functional constraints and non-functional constraints (general constraints). Each functional constraint is made to have $d$ allowed tuples. Except for the functional constraints, all the other constraints are non-functional constraints. We use five main parameters ($n$, $d$, $e$, $nf$, $t$) and six auxiliary parameters (*seed, stpc, stpfc, stpt, stps, instance*) to generate the problems. Table 1 presents both the meanings and values for all the parameters involved in our experiments.

**Table 1. The relevant parameters and their values in the experiments**

| # | Name | Meaning | Value | Start value | End value |
|---|---|---|---|---|---|
| 1 | $n$ | number of variables | 50 | 50 | 50 |
| 2 | $d$ | domain size of each variable | 50 | 50 | 50 |
| 3 | $e$ | number of constraints | 344* | 100 | 710 |
| 4 | $nf$ | number of functional constraints | 2* | 1 | 50 |
| 5 | $t$ | tightness for the constraints | 0.6* | 0.30 | 0.75 |
| 6 | *stpc* | step size for constraints | 122 | 122 | 122 |
| 7 | *stpfc* | step size for functional constraints | 1 | 1 | 1 |
| 8 | *stpt* | step size for tightness | 0.01 | 0.01 | 0.01 |
| 9 | *seed* | seeds | 93728 | 93728 | 11016 |
| 10 | *stps* | step size of seeds | 123 | 123 | 123 |
| 11 | *instance* | number of instances | 1* | 10 | 1 |

In the Table 1, * denotes the first significant value found in the interval [start value, end value]; it may vary a little based on the other parameter settings.

When the problems are generated, we need to select the meaningful instances to construct our data set. In the context of random problems, the generated problem can be easily solved with very few backtrackings, such as 0 backtracking. Also the problem can be exceptionally hard. For example, in our experiments, we discover an exceptionally hard problem instance with $nf = 21$, $e = 710$ and $t = 0.60$, the running time is as much as 560,117.0 seconds with 4.35782e+08 backtrackings.

## 5.1. Data collection

Since the CSPs are NP hard in general, the computational time for the process of keeping on generating and solving all the possible cases can be exceptionally high. First, we attempt many different ways to generate the hard CSPs that are not trivial. After two weeks of random hand-oriented search of hard CSP problems, we finally come up with the idea that the safest way to find all the hard CSP problems is to scan the all possible configurations of the specified problem parameters.

In order to verify the hardness of the CSPs, after having them generated followed by our procedures, we solve them with a general CSP solver, which uses the standard backtracking algorithm. The general solver is implemented in C++ based on the Arc Consistency algorithm (AC 2001/3.1). During the search, the value of a variable is selected in a lexicographical order and the variable with the maximum degree is selected first with the tie broken by lexicographical order.

We use two steps to scan the problems, both with batch files: the first step is to locate the distributions of the hard CSPs in general. Specifically, given that $n$, $d$ are equal to 50, 50 respectively, we use the batch file with the following configurations:
• the number of constraints varying from 100 to 1076 with step size 122, which is about 10% of the total possible constraints
• the number of functional constraint varying from 1 to 50 with step size 1
• the tightness varying from 0.1 to 1.0 with step size 0.01

After about 12 days and nights of running the program on a DELL PowerEdge 1850 (two 3.6 GHz Intel Xeon CPUs) in Linux operating system, we finish the first step of scanning. Based on computational results from the first step scanning, we can safely eliminate the CSP instances with e=100, 222. Furthermore, we can eliminate the CSPs with the tightness intervals of [0.1, 0.30) and (0.75, 0.90]. The second step scanning is to concentrate on the smaller intervals discovered from the first step. Specifically, we scan the interval with following configurations:
• the interval for constraints varying from 344 to 710 with step size 122.
• the interval for tightness varying from 0.30 to 0.75 with step size 0.01.

When systematically examining the problems in our experiments, we maintain the CSP instances with the following settings: $n = 50$ and $d = 50$, which are used widely in the CP community. The detailed values of involved parameters in two steps scanning process are enumerated in Table 1.

## 5.2. Data processing

We classify the CSP instances into three different classes based on the running time. Since the majority of the generated CSPs are trivially solvable with the solving time less than 0.5 second, we mainly consider the CSP instances with the running time more than 60.0 seconds as the hard CSPs. If the solving time is less 5.00 seconds, the instance will be classified as a easy problem. The left instances are classified as the medium problems. Table 2 lists the detailed classification criteria.

**Table 2. Hardness classifications based on the solving time for each CSP**

| Cases # | Time(s) | Classification |
|---|---|---|
| Case 1 | (0-5.00) | Easy |
| Case 2 | [5.00-60.00] | Medium |
| Case 3 | (60.00-inf) | Hard |

Based on the above criteria, we obtain three different classes of problem instance. Since we are only interested in the hard CSPs, we can safely omit the trivial instances. For the rest of the problems, we continue to carry out the association rule mining, which include 266 instances selected from total 10334 instances.

In the stage of data preprocessing, there are two main steps: the first step is to remove the irrelevant data and the noise data such as number of backtrackings. The reason is that for a specified CSP instance, its running time is a monotone increasing function on its backtrackings. The second step is to discretize the data set. In this step, we mainly apply the unsupervised discretization algorithms (simple binning) integrated in WEKA (a collection of data mining algorithms implemented in Java, available at http: //www.cs.waikato.ac.nz/ml/weka/). In this stage, we carry out a set of experiments to prepare the data for the association rule mining. When discretizing the data, we first select the default bin value as 10 for the first time to discretize the item set. In the following, to balance the accuracy and the redundancy, we eventually choose to set the bin value as 20 to discover the interesting association rules from the selected data set.

In the stage of association rule mining, we also conduct a set of experiments to discover the interesting rules. Table 3 presents top 10 interesting association rules generated by Apriori algorithm on the data set discretized with the bin value 20. The selected measure is lift and its value is set to 1.1 while the minimum support is set to 0.1. The rules in the table 3 are listed in decreasing order of the lift. As a matter of fact, since we use lift as the main measure, WEKA always find the symmetric rule simultaneously (i.e., given the specified thresholds, if the rule X $\Longrightarrow$ Y is generated, then its symmetric rule Y $\Longrightarrow$ X follows it immediately). Table 3 actually indicates 20 different association rules. For example, rule 1 corresponds to its symmetric rule:

SAT=NO Classification=Hard $\Longrightarrow$ e=(691.7-710] t=(0.6075-0.628]

The above rule and rule 1 have the same lift value 1.56, which is guaranteed by the definition of the lift (an even function). For purpose of simplifying the problem representation, we only present 10 rules in table 3.

From table 3, we can clearly observe the measures and interesting rules. For example, rule 4 indicates that if the total constraints of a CSP is more than 692 and their tightness are located in the interval of (0.6075-0.628], then this CSP is highly likely to be a hard CSP instance. The lift, confidence and conviction of this rule are 1.39, 0.55 and 1.29 respectively.

Rule 1, 3, 4, 5, 6, 7, 10 in table 3 all indicate that the hard CSPs are usually accompanied with more than 692 constraints on the basis of n=50 and d=50.

### 5.3. Interesting results

From the mined association rules in Table 3, we can see that the hard CSPs are highly like to be unsatisfiable, which make the backtracking number maximum for the given CSP. The conclusion is consistent with previous conclusions made in [4].

An interesting phenomenon indicated from rule 9 of table 3 is that the tightness of the hard CSPs usually lies in the interval of (0.6075, 0.628], which is near the golden means (0.618). Furthermore, when the total constraints are bigger than 692 and less than 710, the problem are likely to be hard. (Note these rules are generated from the medium hard and the hard CSP instances whose number of variables and their domain size are set to 50 and 50 respectively).

By applying the rule deduction, especially transitivity and augmentation axioms, on the total 62 interesting rules generated by WEKA, we eventually obtain the interesting patterns of hard CSPs as shown in Figure 3. This figure presents the CSP hardness patterns with the hard CSP located at the center and other important characteristics around. The pattern reveals the potential relationship among the CSPs hardness, tightness of constraints, number of functional constraints (if any), and total constraints.

**Table 3. The best ten association rules and their interesting measures (SAT=NO denotes that the CSP instance is unsatisfiable; SAT=YES denotes that it is satisfiable)**

| # | Interesting association rules | Interestingness measures | | |
|---|---|---|---|---|
| | | Lift | Confidence | Conviction |
| 1 | e=(691.7-710] t=(0.6075-0.628] => SAT=NO Classification=Hard | 1.56 | 0.42 | 1.23 |
| 2 | nf=(37.75-40.2] => Classification=Medium | 1.54 | 0.93 | 3.82 |
| 3 | t=(0.6075-0.628] => e=(691.7-710] SAT=NO Classification=Hard | 1.40 | 0.35 | 1.13 |
| 4 | e=(691.7-710] t=(0.6075-0.628] => Classification=Hard | 1.39 | 0.55 | 1.29 |
| 5 | e=(691.7-710] t=(0.6075-0.628] SAT=NO => Classification=Hard | 1.34 | 0.53 | 1.23 |
| 6 | t=(0.6075-0.628] SAT=NO => e=(691.7-710] Classification=Hard | 1.30 | 0.47 | 1.17 |
| 7 | nf=(27.95-30.4] => e=(691.7-710] SAT=NO | 1.30 | 0.75 | 1.54 |
| 8 | e=(581.9-600.2] => Classification=Medium | 1.28 | 0.78 | 1.58 |
| 9 | t=(0.6075-0.628] => SAT=NO Classification=Hard | 1.28 | 0.35 | 1.09 |
| 10 | e=(691.7-710] SAT=NO => t=(0.6075-0.628] Classification=Hard | 1.26 | 0.18 | 1.04 |

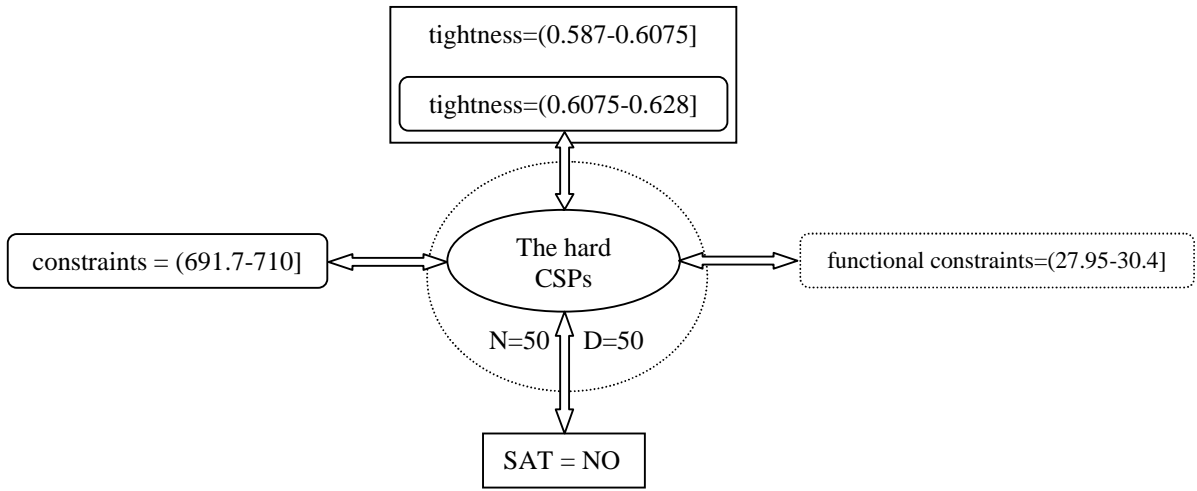

**Figure 3. The interesting patterns of the randomly generated hard CSPs with both number of variables and domain set to 50**

In Figure 3, the tightness has two intervals, both of which satisfy the specified thresholds on support and lift. The difference is that the first tightness interval (0.6075, 0.628] can generate the hard CSP instances with higher probability than the interval of (0.587, 0.6075]. The CSP hardness patterns reveal the close associations among number of constraints and number of functional constraints and their tightness, when these parameters can generate the hard CSPs. If one want to generate the hard CSP instances with the parameters ($n, d, e, tightness$), this CSP hardness pattern can provide a very good recommendations. Simply set ($n = 50$, $d = 50$, $e = 700$, $tightness = 0.618$, functional constraints = 29 (if any)) and the hard CSPs will be generated.

## 6. Discussion and remarks

In terms of searching hard CSPs, although the safest way is to scan all the possible CSP instances and solve them completely, it only holds for the ideal status because there are innumerous problem instances, which may take the running time last forever. In our experiments, we use 10 parameters (i.e., $n$, $d$, $e$, $nf$, tightness, step size for e, $nf$, $t$, seed and its step size) to control the scanning process of the generated CSP instances. Even for the CSPs with the fixed number of variables and domain size (in our experiments, we set $n = 50$, $d = 50$), the combination of the rest parameters can be extremely huge. Suppose we make $e$ varying from 100 to 1100 with step size 1, $nf$ from 1 to 100 with step size 1, tightness from 0.01 to 0.99 with step size 0.01, and seed from 93728 to 96728 with step size 3, the problem instance number will be as many as $10^{10}$. If solving each instance takes roughly 10.00 seconds on average, the total running time would be about 106 years. Besides, the step size can be significantly small, say $10^{-3}$ or $10^{-6}$, based the size of the variable domains. If the number of variables and domain size are increasing as well, the problem instances will increase exponentially. Therefore, back to our experiments, we first scan the CSP distribution space sparsely to obtain the typical distribution of hard CSPs; then we concentrate on scanning in the relatively small distribution space established at the first step scanning.

For the randomly generated CSPs, the hard CSP are usually characterized by the tightness located in the interval of (0.6075, 0.628]. For the researchers who want to generate the hard CSPs with functional constraints, based on the patterns of hard CSPs, we recommend them to set the constraints tightness near the golden means (0.618), and the functional constraint 28, 29 or 30, which is about 4.3% of the generated constraints. The total constraints in the generated CSP are roughly 60.0% of all the possible constraints in the whole constraint network.

## 7. Conclusion

Hard CSPs are highly evaluated for their wide usability in the CP community. In this paper, for the first time, the patterns of the hardness of CSPs are investigated with the association rule mining techniques. With our method, we generate the CSPs with the parameter set <*n*, *d*, *e*, *nf*, *t*, *ft*, *seed*>. We study the experimental data within the binary CSPs with functional constraints, bi-functional constraints and non functional constraints. In order to verify the hardness of the generated CSPs, we solve the problem instances with a general CSP solver implemented based on the AC2001/3.1. We systematically scan the distributions space of the CSPs based on the specified parameters. By solving all the instances, we obtain the important characteristics of the CSPs as the raw data set. Then we classified the data according to the running time of each instance. Following that we apply the association rule mining techniques to the selected data set. Finally, we successfully discover the CSP hardness patterns under the rule deduction scheme.

## 8. Future work

One of the significant directions for the future work is to apply the extracted CSPs hardness patterns to guide the searching process for hard CSPs. With the help of CPS hardness patterns, we can precisely locate the intervals of all the relevant parameters. Moreover, we can discover more hard CSPs and much harder CSPs to benchmark the CSP solvers, even SAT solvers.

Another interesting direction is to integrate the resolution techniques from propositional logic with the association rules to generate more interesting information on the CSP hardness patterns. In that case, more accurate guidance can be provided for searching the hard CSPs.

## 9. Acknowledgements

The author would like to thank Yichen Liu her useful suggestions on the paper writing and her help on the proofreading of the paper draft.